\documentclass[aps,reprint,prb]{revtex4-1}
\usepackage{graphicx}
\usepackage{bm}
\begin{document}
\title{Intraband resonance Raman scattering in anisotropic quantum dots}
\author{A.V.~Shorokhov}
\email{shorokhovav@math.mrsu.ru} \affiliation{Mordovian State University, 430005 Saransk, Russia}
\author{V.A.~Margulis}
\begin{abstract}
We have developed a theory of the one-phonon intraband resonance Raman scattering (IRRS) in anisotropic quantum dots
subjected to an arbitrarily directed magnetic field. The differential Raman cross section is obtained. The resonance
structure of the Raman cross section is studied. It is shown that the quantum dot subjected in a magnetic field can be
used as the detector of phonon modes. The interesting multiplet structure of the resonance peaks is studied.
\end{abstract}
\pacs{78.67.Hc, 78.30.Fs}
\maketitle
\section{Introduction}
Theoretical \cite{Fomin98,Menendez99,Klimin08,Rodr00,Chamb95,Fedorov97,Men02} and experimental
\cite{Heitz00,Milekhin04,Choi05,Sarkar08,Sotom08} studies of the Raman scattering and photoluminescence in quantum dots
(QD) are of great interest because the understanding of the scattering mechanisms is of fundamental importance for the
applications. Raman scattering can provide the direct information on the electronic structure, phonon spectrum, and
optical properties of QDs \cite{Cardona89}. The most part of papers is devoted to studying the {\it interband} Raman
scattering. However, we suppose that the {\it intraband} resonance Raman scattering is also of great interest because
the distance between discrete levels in QDs can be done of order the optical phonon energy with help of the magnetic
field. It is important to note that the optical phonon emission is known to play a dominant role in QDs among the
scattering mechanisms present in polar semiconductors.

Modern nanotechnology enables one to fabricate quantum dots of different shapes. In the past years the significant
interest has been given to quantum wells and QDs that are characterized by an asymmetric confining potential
\cite{Ahn87,Zhang05,Yild06}. In this work we present a theoretical study of the intraband resonance Raman scattering of
an anisotropic quantum dot subjected to a uniform magnetic field arbitrarily directed with respect to the potential
symmetry axes. The applied magnetic field gives us the possibility to change the distance between levels and to adjust
the energy levels of QDs on the various phonon modes. The study of the different polarization for the incident and
emitted radiation yields the additional information about the phonon spectrum. Note that the study of intraband Raman
scattering lets us to obtain the simple analytic formulae for the cross-section in the case of anisotropic QDs.

\begin{figure}
\includegraphics[clip=true,width=0.7\linewidth]{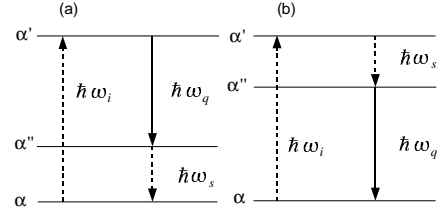}
\caption{\label{fig1} Transitions leading to resonant absorption.}
\end{figure}

Resonance intraband Raman scattering in our case can be qualitatively described in the following way: the absorption of
quantum $\hbar\omega_i$ of the high-frequency field (laser pump), emission of optical phonon $\hbar\omega_{\mathbf{q}}$
(photon $\hbar\omega_s$) in an intermediate state and emission of photon $\hbar\omega_s$ (optical phonon
$\hbar\omega_{\mathbf{q}}$) in the initial state (Figure \ref{fig1}). In this approach, the Raman cross section is
calculated from third order time-dependent perturbation theory.

We model the semiconductor QD using the asymmetric parabolic confinement $V({\mathrm
r})=m^*(\Omega_x^2x^2+\Omega_y^2y^2+\Omega_z^2z^2)/2$ (here $m^*$ is the electron effective mass, $\Omega_i$
($i=x,y,z$) are the characteristic frequencies of the parabolic potential). The spectrum of electrons in such dot
placed in an arbitrarily directed magnetic field $\mathbf{B}$ with the vector potential
$\mathbf{A}=(B_yz/2-B_zy,0,B_xy-B_yx/2)$ has the form
 $\varepsilon_{nml}=\hbar\omega_1(n+1/2)+\hbar\omega_2(m+1/2)+\hbar\omega_3(l+1/2)$ ($n,m,l=0,1,2,\ldots$), where hybrid frequencies
 $\omega_j$ ($j=1,2,3$) are obtained from the sixth-order algebraic equation \cite{Geyler05}.

\section{Raman cross section}
The differential resonance Raman cross section $d^2\sigma/d\Omega d\omega_s$ of a volume $V$ per unit solid angle
$d\Omega$ for incident radiation with the frequency $\omega_i$ and emitted radiation with the frequency $\omega_s$ is
given by \cite{Chamb95}
\begin{equation}
\frac{d^2\sigma}{d\Omega d\omega_s}=\frac{V^2\omega_s^3n_in_s^3}{8\pi^3c^4\omega_i}W(\omega_s,{\mathbf e}_s)
\end{equation}
where $n_i(n_s)$ is the refractive index of the medium with frequency $\omega_i$ ($\omega_s$), $c$ is the velocity of
light, ${\mathbf e}_s$ is the unit polarization vector of the emitted radiation. The transition rate is calculated
according to
\begin{equation}
W(\omega_s,{\mathbf e}_s)=\frac{2\pi}{\hbar}\sum_\alpha
\left|W_{\alpha\alpha}\right|^2\delta(\hbar\omega_i-\hbar\omega_s-\hbar\omega_q),
\end{equation}
where $\alpha=(n,m,l)$ are the quantum numbers of the initial electron states in QD.

We consider only resonance transitions. In this case  the scattering amplitude probability for phonon emission in QDs
is described by a sum of two terms
\begin{eqnarray}
\label{amp1} &&W_{\alpha\alpha}=\sum_{\alpha',\alpha''}\frac{\langle\alpha|\hat{H}_R(\omega_s)|\alpha''\rangle
\langle\alpha''|\hat{H}_L|\alpha'\rangle\langle\alpha'|\hat{H}_R(\omega_i)|\alpha\rangle}
 {(\varepsilon_{\alpha'}-\varepsilon_{\alpha}-\hbar\omega_i)
 (\varepsilon_{\alpha''}-\varepsilon_{\alpha}-\hbar\omega_i+\hbar\omega_{\mathbf q})}\nonumber\\
&+&\sum_{\alpha',\alpha''}\frac{\langle\alpha|\hat{H}_L|\alpha''\rangle
\langle\alpha''|\hat{H}_R(\omega_s)|\alpha'\rangle\langle\alpha'|\hat{H}_R(\omega_i)|\alpha\rangle}
 {(\varepsilon_{\alpha'}-\varepsilon_{\alpha}-\hbar\omega_i)
 (\varepsilon_{\alpha''}-\varepsilon_{\alpha}-\hbar\omega_i+\hbar\omega_s)}
\end{eqnarray}
The first term in Eq.(\ref{amp1}) corresponds to the transitions depicted on Figure \ref{fig1}a, the second term in Eq.
(\ref{amp1}) corresponds to the transitions depicted on Figure \ref{fig1}b.

Here $H_L$ is the operator electron-phonon interaction
\begin{equation}
{\hat H}_L=\sum_{\mathbf{q}}D_{\mathbf{q}}C_\mathbf{q}\exp(i\mathbf{q}\mathbf{r})+\mathrm{c.c},
\end{equation}
where $D_{\mathbf{q}}$ is the electron-phonon coupling constant.

The operator of the electron-photon interaction can be expressed as
\begin{equation}
\hat{H}_R=\frac{e}{m^*}\sqrt{\frac{2\pi\hbar N}{\varepsilon\omega}}\mathbf{e}\mathbf{P},
\end{equation}
where $N$ is the number of initial-state photons with frequency $\omega$, $\mathbf{e}$ is the polarization vector and
$\mathbf{P}=\mathbf{p}-e\mathbf{A}/c$ is the generalized momentum, $\varepsilon$ is the real part of the dielectric
constant.

 A direct calculation of the matrix elements of electron-photon and electron-phonon
 interactions is a complicate problem in our case. However, the method of canonic transformation of the phase
space allows us to resolve this problem using only simple calculations from linear algebra \cite{Geyler05}. In
particular, in our preceding papers we used this method to study hybrid \cite{Geyler01}, hybrid-phonon
\cite{Margulis09} and hybrid-impurity resonances in this system \cite{Margulis02}.

Using the results obtained in \cite{Geyler01}, we can write the matrix elements of the operator $\hat{H}_R$
 in the following form
\begin{eqnarray}
&&\langle n'm'l'|\hat{H}_R|nml\rangle=ie\hbar\sqrt{\frac{\pi N}{m^*\varepsilon\omega}}\nonumber\\
&\times&\left[X_1\sqrt{n+1}\delta_{n', n+1 }\delta_{m', m}\delta_{l', l}\right.\nonumber\\
&+&X_2\sqrt{m +1}\delta_{n,n'}\delta_{m',m+1}\delta_{l',l}\nonumber\\
&+&\left.X_3\sqrt{l+1}\delta_{n',n}\delta_{m',m}\delta_{l',l+1}\right]. \label{HR}
\end{eqnarray}
where the coefficients $X_i$ ($i=1,2,3$) were found in \cite{Geyler01}.

We introduce the notation
\begin{eqnarray}
&&J(n''m''l'',n'm'l')=\left(\frac{n''!m''!l''!}{n'!m'!l'!}\right)^{1/2}(-1)^{n'-n''}\nonumber\\
&\times&(-1)^{m'-m''}(-1)^{l'-l''}
\exp[i\varphi_1(n'-n'')]\nonumber\\
&&\times\exp[i\varphi_2(m'-m'')]\exp[i\varphi_3(l'-l'')]g^{n'-n''}_{1}g^{m'-m''}_{2}\nonumber\\
&&\times g^{l'-l''}_{3}L^{n'-n''}_{n''}(g^{2}_1)L^{m'-m''}_{m''}(g^{2}_2) L^{l'-l''}_{l''}(g^{2}_3). \label{J}
\end{eqnarray}
Here $g_j=\sqrt{\lambda_j^2+\kappa_j^2l_j^4}/\sqrt{2}l_j$, $\tan\varphi_j=\kappa_j l_j^2/\lambda_j$,
$l_j=\sqrt{\hbar/m^*\omega_j}$ ($j=1,2,3$) are the hybrid lengths, , $L^{n'}_{n}(x)$ are the generalized Laguerre
polynomials, $\lambda_i=\hbar(b_{1i}q_x+b_{2i}q_y+b_{3i}q_z)$ ($i=1,2,3$), $\kappa_{i-3}=b_{1i}q_x+b_{2i}q_y+b_{3i}q_z$
($i=4,5,6$), $b_{ji}$ are components of the transition matrix \cite{Margulis02}.

Using (\ref{J}), we can write the matrix elements of the operator electron-phonon interaction as
\begin{eqnarray}
&&\label{HL}\langle n'm'l'|\hat{H}_L|n''m''l''\rangle=\sum_{\mathbf{q}}D_q\sqrt{N_\mathbf{q}}\exp(-g^2/2)\nonumber\\
&\times&\exp[-(\kappa_1\lambda_1+\kappa_2\lambda_2+\kappa_3\lambda_3)i/2]J(n''m''l'',n'm'l'),
\end{eqnarray}
where $N_\mathbf{q}$ is the number of phonons with the wave vector $\mathbf{q}$ and $g^2=g_1^2+g_2^2+g_3^2$.

Substituting (\ref{HR}) and (\ref{HL}) into (\ref{amp1}) after some cumbersome algebra it is possible to get analytic
expression for $W_{\alpha\alpha}$. We consider only the resonance Raman scattering. In this case the frequency of the
pump is equal to the distance between the levels of QD. For definiteness, assume that we pump the QD by the laser with
frequency $\omega_i=\omega_1$. Then we need to keep only resonance terms in the formula for $W_{\alpha\alpha}$. In this
case for the first term in Eq. (\ref{amp1}), we obtain
\begin{eqnarray}
\label{a} &&\sum_{\alpha',\alpha''}\frac{\langle\alpha|\hat{H}_R(\omega_s)|\alpha''\rangle
\langle\alpha''|\hat{H}_L|\alpha'\rangle\langle\alpha'|\hat{H}_R(\omega_i)|\alpha\rangle}
 {(\varepsilon_{\alpha'}-\varepsilon_{\alpha}-\hbar\omega_i)
 (\varepsilon_{\alpha''}-\varepsilon_{\alpha}-\hbar\omega_i+\hbar\omega_{\mathbf q})}\nonumber\\
 &&=-\frac{\pi e^2}{m^*\varepsilon}\sqrt{\frac{N_i(N_s+1)}{\omega_i\omega_s}}
 \sum_{\mathbf{q}}D_q\sqrt{N_\mathbf{q}}\exp(-g^2/2)\nonumber\\
&\times&\exp[-(\kappa_1\lambda_1+\kappa_2\lambda_2+\kappa_3\lambda_3)i/2]
\frac{\sqrt{n+1}X_1^i}{\omega_1-\omega_i}\nonumber\\
&\times&\left[\frac{\sqrt{m+1}X_2^SJ(nm+1l,n+1ml)}{\omega_2-\omega_i+\omega_\mathbf{q}}\right.\nonumber\\
&&\left.+
\frac{\sqrt{l+1}X_3^SJ(nml+1,n+1ml)}{\omega_3-\omega_i+\omega_\mathbf{q}} \right]. \nonumber\\
\end{eqnarray}
Here the indexes $i$ and $s$ are referred to the incident and emitted photons, respectively.

The second term in (\ref{amp1}) has the following form
\begin{eqnarray}
\label{amp} &&=\sum_{\alpha',\alpha''}\frac{\langle\alpha|\hat{H}_L|\alpha''\rangle
\langle\alpha''|\hat{H}_R(\omega_s)|\alpha'\rangle\langle\alpha'|\hat{H}_R(\omega_i)|\alpha\rangle}
 {(\varepsilon_{\alpha'}-\varepsilon_{\alpha}-\hbar\omega_i)
 (\varepsilon_{\alpha''}-\varepsilon_{\alpha}-\hbar\omega_i+\hbar\omega_s)}\nonumber\\
 &&=
 -\frac{\pi e^2}{m^*\varepsilon}\sqrt{\frac{N_i(N_s+1)}{\omega_i\omega_s}}\sum_{\mathbf{q}}D_q\sqrt{N_\mathbf{q}}\exp(-g^2/2)\nonumber\\
&\times&\exp[-(\kappa_1\lambda_1+\kappa_2\lambda_2+\kappa_3\lambda_3)i/2]
\frac{\sqrt{n+1}X_1^i}{\omega_1-\omega_i}\nonumber\\
&\times&\left[\frac{\sqrt{m}X_2^SJ(nml,n+1m-1l)}{\omega_1-\omega_2-\omega_i+\omega_s}\right.\nonumber\\
&&\left.+ \frac{\sqrt{l}X_3^SJ(nml,n+1ml-1)}{\omega_1-\omega_3-\omega_i+\omega_s} \right]. \label{b}
\end{eqnarray}
Now we need to sum these terms to get $W_{\alpha\alpha}$. Taking into account the conversation law
$\hbar\omega_i=\hbar\omega_\mathbf{q}+\hbar\omega_s$ we can transform denominators in Eq. (\ref{a}) and Eq. (\ref{b}).
In this case we get for $W_{\alpha\alpha}$ the following formula
\begin{eqnarray}
\label{W} W_{\alpha\alpha}&=&-\frac{\pi e^2}{m^*\varepsilon}\sqrt{\frac{N_i(N_s+1)}{\omega_i\omega_s}}
\sum_{\mathbf{q}}D_q\sqrt{N_\mathbf{q}}\exp(-g^2/2)
\nonumber\\
&\times&\exp[-(\kappa_1\lambda_1+\kappa_2\lambda_2+\kappa_3\lambda_3)i/2]\frac{\sqrt{n+1}X_1^i}{\omega_1-\omega_i}\nonumber\\
&\times&\left\{\frac{X_2^S}{\omega_2-\omega_s}\left[\sqrt{m+1}J(nm+1l,n+1ml)\right.\right.\nonumber\\
&&\left.\left.-\sqrt{m}J(nml,n+1m-1l)\right]\right.
\nonumber\\
&+&\left.\frac{X_3^S}{\omega_3-\omega_s}\left[\sqrt{l+1}J(nml+1,n+1ml)\right.\right.\nonumber\\
&&\left.\left.-\sqrt{l}J(nml,n+1ml-1)\right]\right\}\nonumber\\
\end{eqnarray}
We can transform the differences  in Equation (\ref{W}) using the recurrent formula for the Laguerre polynomials
$xL_n^{\alpha+1}=(n+\alpha+1)L_n^{\alpha}(x)-(n+1)L_{n+1}^{\alpha}(x)$. As a result we get for the first difference
\begin{eqnarray}
&&\sqrt{m+1}J(nm+1l,n+1ml)-\sqrt{m}J(nml,n+1m-1l)\nonumber\\
&=&-\exp(i\varphi_1)\exp(-i\varphi_2)g_1g_2L_n^1(g_1^2)L_m(g_2^2)L_l(g_3^2)\label{dif}
\end{eqnarray}
The second difference in Eq. (\ref{W}) is calculated in analogy with Eq. (\ref{dif}). Using the obtained estimation, we
get the following formula for the square of the scattering amplitude probability
\begin{eqnarray}
&&|W_{\alpha\alpha}|^2=\frac{\pi^2e^4}{m^{*2}}\frac{N_i(N_s+1)}{\omega_i\omega_s}
\sum_{\mathbf{q}}|D_\mathbf{q}|^2N_\mathbf{q}\exp(-g^2)g_1^2\nonumber\\
&&\times[L_n^1(g_1^2)]^2[L_m(g_2^2)]^2[L_l(g_3^2)]^2\frac{(n+1)(X_1^i)^2}{(\omega_1-\omega_i)^2}\nonumber\\
&&\times
\left|\frac{g_2X_2^S}{\omega_2-\omega_s}\exp(-i\varphi_2)+\frac{g_3X_3^S}{\omega_3-\omega_s}\exp(-i\varphi_3)\right|^2.
\end{eqnarray}

Then we can write the final formula for the Raman cross-section taking into account the smearing of the hybrid levels
by collisions
\begin{eqnarray}
\label{sigma} &&\frac{d^2\sigma}{d\Omega d\omega_s}=\frac{V\omega_s^2n_ie^4N_i(N_s+1)}{4\hbar^2 n_sc^4m^{*2}}
\sum_{\mathbf{q}}|D_\mathbf{q}|^2N_\mathbf{q}\exp(-g^2)
\nonumber\\
&\times&g_1^2[L_n^1(g_1^2)]^2[L_m(g_2^2)]^2[L_l(g_3^2)]^2 \frac{(n+1)(X_1^i)^2}{(\omega_1-\omega_i)^2+\Gamma^2}\nonumber\\
 &\times&\left|\frac{g_2X_2^S}{\omega_2-\omega_s-i\Gamma}\exp(-i\varphi_2)+\frac{g_3X_3^S}{\omega_3-\omega_s-i\Gamma}
\exp(-i\varphi_3)\right|^2\nonumber\\
&&\times \delta(\omega_i-\omega_s-\omega_q),
\end{eqnarray}
where $\Gamma$ is the lifetime broadening.
\section{Results and discussions}

\begin{figure}
\includegraphics[clip=true,width=0.7\linewidth]{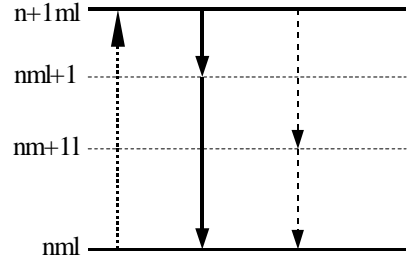}
\caption{\label{fig2} Possible transitions leading to the resonant absorption in the case of anisotropic quantum dots.
The dotted curve corresponds to the absorption of the pump field with the frequency $\omega_1$. The solid line
corresponds to the transitions with the change of the quantum number $m$. The dashed curve corresponds to the
transitions with the change of the quantum number $l$.}
\end{figure}

Equation (\ref{sigma}) clearly shows that if one ignores the optical phonons dispersion and if the frequency of the
pump is equal to $\omega_1$ then we have the input resonance on the frequency $\omega_1$ and the output resonance on
the frequencies $\omega_2$ and $\omega_3$. Note that it is forbidden transitions with simultaneous changing more than
one quantum numbers in the case of absorption (emission) of photon due to the selection rules. The possible transitions
is shown on the Figure \ref{fig2}.

It is important to note that the hybrid frequency $\omega_k$ ($k=1,2,3$) is determined by the magnitude and the
direction of the magnetic field (i.e. they can be tuned with the help of the magnetic field). Hence, using the tunable
laser and changing, for example, the magnitude of the magnetic field we can register phonon modes (with frequencies
$\omega_q=\omega_1-\omega_2$ and $\omega_q=\omega_1-\omega_3$) in quantum dots as series resonance peaks in the
dependence of the Raman cross section on the magnetic field. The frequency of the phonon mode can be determined from
the dependence of the magnetic field on the hybrid frequencies.

Let us now to study effects arising due to the dispersion of the phonons. Replacing the sum over the phonon wave vector
by the integral in Equation (\ref{sigma}) and assuming a parabolic dispersion low for long-wave phonons
$\omega_q=\omega_0(1-\omega_0^{-2}V_s^2q^2)$, where $\omega_0$ is the optical-phonon threshold frequency and $V_s$ is
the speed of sound, we can easily evaluate the integral with respect to $|\mathbf{q}|$ thanks to the presence of a
delta function $\delta(\omega_i-\omega_s-\omega_q)$.
\begin{figure}
\includegraphics[clip=true,width=0.8\linewidth]{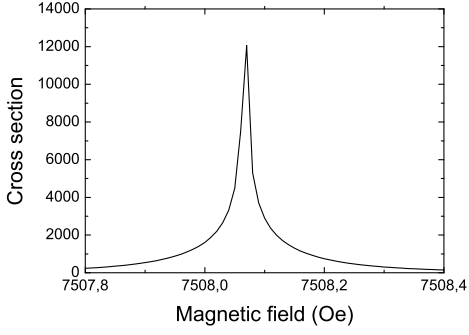}
\caption{\label{fig3} Raman cross section (in arb. units) as a function of a magnetic field in the case of transition
from the ground state and emission of PO-phonons, $\omega_0=1.2\times10^{12}$ s$^{-1}$, $\omega_x=1.3\times10^{12}$
s$^{-1}$, $\omega_y=1.4\times10^{12}$ s$^{-1}$, $\omega_z=7.1\times10^{13}$ s$^{-1}$.}
\end{figure}

Converting to spherical coordinates we obtain the following equation for the Raman cross section
\begin{eqnarray}
\label{sigma1} &&\frac{d^2\sigma}{d\Omega
d\omega_s}=\frac{V\omega_s^2n_ie^4N_i(N_s+1)N_0\omega_0^{3/2}}{8\hbar^2V_s^3n_s c^4m^{*2}\sqrt{|\Delta\omega_0|}}
\nonumber\\
&&\times\int\limits_{0}^{2\pi}d\varphi\int\limits_{0}^{\pi}\sin\theta d\theta|D_\mathbf{q}|^2\exp(-y^2)y_1^2\nonumber\\
&&\times [L_n^1(y_1^2)]^2[L_m(y_2^2)]^2[L_l(y_3^2)]^2\frac{(n+1)(X_1^i)^2}{(\omega_1-\omega_i)^2+\Gamma^2}\nonumber\\
&&\times\left|\frac{y_2X_2^S}{\omega_2-\omega_s-i\Gamma}\exp(-i\varphi_2)+\frac{y_3X_3^S}{\omega_3-\omega_s-i\Gamma}
\exp(-i\varphi_3)\right|^2,\nonumber\\
\end{eqnarray}
Here we replace $N_{\mathbf{q}}$ by the Plank distribution function $N_0$, $y_j$ can be obtained from $g_j$ if we write
the vector $\mathbf{q}$ in the spherical coordinates, $D_{\mathbf{q}}$ depends on the electron-phonon interaction and
$\Delta\omega_0=\omega_1-\omega_s-\omega_0$.

Let us consider, first of all, the polarization potential scattering (PO phonons). In this case the electron-phonon
coupling constant
\begin{equation}
|D_{\mathbf{q}}|^2=\frac{2\sqrt{2}\pi\hbar^2\alpha_L\omega_0^{3/2}}{\sqrt{m^*}q^2}.
\end{equation}
 Then we can rewrite Equation (\ref{sigma1})
as
\begin{eqnarray}
\label{sigma2} &&\frac{d^2\sigma}{d\Omega d\omega_s}=\frac{\sqrt{2}\pi
V\omega_s^2n_in_s^3e^4N_i(N_s+1)N_0\omega_0^2}{4V_s^3 c^4m^{*3/2}\varepsilon^2|\Delta\omega_0|^{3/2}}\nonumber\\
&&\int\limits_{0}^{2\pi}d\varphi\int\limits_{0}^{\pi}\sin\theta d\theta|D_\mathbf{q}|^2\exp(-y^2)y_1^2
\nonumber\\
&&\times[L_n^1(y_1^2)]^2[L_m(y_2^2)]^2[L_l(y_3^2)]^2 \frac{(n+1)(X_1^i)^2}{(\omega_1-\omega_i)^2+\Gamma^2}\nonumber\\
&&\times\left|\frac{y_2X_2^S}{\omega_2-\omega_s-i\Gamma}\exp(-i\varphi_2)+\frac{y_3X_3^S}{\omega_3-\omega_s-i\Gamma}
\exp(-i\varphi_3)\right|^2,\nonumber\\
\end{eqnarray}
The Raman cross section depends on the polarization both the input signal and output one. Let us consider the case when
the polarization vector of the incident and emitted fields are perpendicular to the magnetic field. In this case the
hybrid frequencies are determined by the formulae $\omega_{1,2}=[\sqrt{(\Omega_x+\Omega_y)^2+\omega_c^2}\pm
\sqrt{(\Omega_x-\Omega_y)^2+\omega_c^2}]/2$, $\omega_3=\Omega_z$, and the values of $y_j$ ($j=1,2,3$) have the
following form
\begin{eqnarray}
&&y_j=\frac{l_j}{\sqrt{2}}\sqrt{\frac{\omega_0|\Delta\omega_0|}{V_s^2}}\frac{\sin\theta}{\sqrt{\omega_c^2\Omega_x^2+
(\Omega_x^2-\omega_j^2)^2}}\nonumber\\
&&\times\left[(\Omega_x^2-\omega_i^2)^2\sin^2\varphi+\omega_c^2\omega_i^2\cos^2\varphi\right]^{1/2}
\nonumber\\
\end{eqnarray}
Note that in this case $X_j^i=X_j^S$ ($j=1,2,3$). Equation (\ref{sigma2}) clearly shows that the Raman cross section
has singularities at the points where $\Delta\omega_0=0$. On the Figure \ref{fig3} it is shown the dependence of the
Raman cross section on the magnetic field (here we taken into account the finite phonon relaxation time).
\begin{figure}
\includegraphics[clip=true,width=0.8\linewidth]{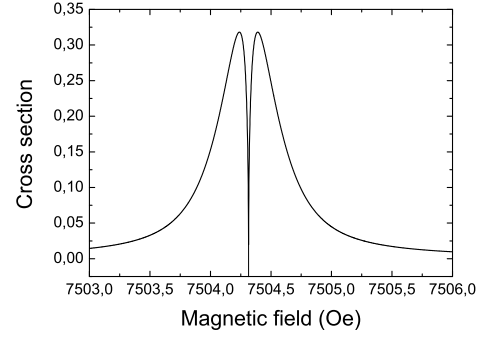}
\caption{\label{fig4} Raman cross section (in arb. units) as a function of a magnetic field in the case of transition
from the ground state and emission of DO-phonons, Other parameters coincide with those of Fig. \ref{fig3}}
\end{figure}

The different situation takes place in the case of deformation potential scattering (DO phonons). The Raman cross
section of deformation potential scattering connected with one of polarization potential scattering by the following
estimation
\begin{eqnarray}
\frac{d^2\sigma_{PO}}{d\Omega d\omega_s}=\frac{mV_s^2}{2\hbar|\Delta\omega_0|}\frac{d^2\sigma_{DO}}{d\Omega d\omega_s}
\end{eqnarray}
\begin{figure}
\includegraphics[clip=true,width=0.8\linewidth]{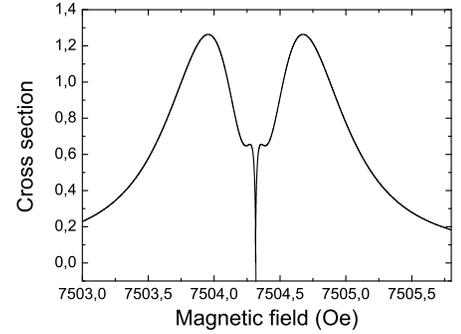}
\caption{\label{fig4} Raman cross section (in arb. units) as a function of a magnetic field in the case of transition
from the state $n=m=l=1$ and emission of DO-phonons. Other parameters coincide with those of Fig. \ref{fig3}}
\end{figure}
It is important to note that in the points where $\Delta\omega_0=0$ the Raman cross section is equal to zero in
contradiction to the case of PO-phonons. In the case of DO-phonons the cross section has the complex doublet structure.
The width of the resonance curve is enough small (of order $1$ Oe) in this situation. In the most simple case of
transitions from the ground state $n=m=l=0$ the resonance curve consists of two symmetrically positioned sharp peaks to
the left and right of the point $\Delta\omega_0=0$ (Fig.4). In the case of transitions from the state $n=m=l=1$ each of
the doublet peaks splits up into two ones (Fig.5). If we take into account the finite phonon relaxation time in QDs,
the resonance curve doesn't change its form in the difference from the polarization scattering but its minimum
displaces in the point where $\Delta\omega+\tau^{-1}=0$ (here $\tau$ is the relaxation time).

In conclusion, we have investigated theoretically the resonance Raman scattering in the anisotropic quantum dots in the
presence of arbitrarily directed magnetic field and polarization vector. Raman scattering lets us to detect phonon
modes in QD using the tunable laser and changing magnetic field. If we ignore optical phonon dispersion, we have a
resonance peak corresponding to the emission of optical phonon mode. The interesting doublet structure of peaks arises
if one takes into account the dispersion of long-wave optical phonons in the case of deformation scattering. In this
case the resonances let us to observe the threshold frequency of optical phonons. In the resonance point the cross
section is equal to zero but in a small neighborhood of this point cross section has symmetrically positioned (to the
left and right) peaks. The number of peaks  depends on the initial quantum state. We hope that our calculations can
further stimulate more experimental measurements on Raman intensities of semiconductor QDs.

\begin{acknowledgments} Present work was supported by the Russian Foundation for Basic Research and the Grant of President
of Russia for Young Scientists (MK-2062.2008.2).
\end{acknowledgments}

\end{document}